# Disordered, quasicrystalline and crystalline phases of densely packed tetrahedra

Amir Haji-Akbari[1]*, Michael Engel[1]*, Aaron S. Keys[1], Xiaoyu Zheng[3], Rolfe G. Petschek[5], Peter Palffy-Muhoray[4] & Sharon C. Glotzer[1,2]

[1]Department of Chemical Engineering, [2]Department of Materials Science and Engineering, University of Michigan, Ann Arbor Michigan 48109, USA, [3]Department of Mathematical Sciences, [4]Liquid Crystal Institute, Kent State University, Kent, Ohio 44242, USA, [5]Department of Physics, Case Western Reserve University, Cleveland Ohio 44106, USA.
*These authors contributed equally to this work.

**All hard, convex shapes are conjectured by Ulam to pack more densely than spheres [1], which have a maximum packing fraction of $\phi = \pi/\sqrt{18} \approx 0.7405$. For many shapes, simple lattice packings easily surpass this packing fraction [2][3]. For regular tetrahedra, this conjecture was shown to be true only very recently; an ordered arrangement was obtained via geometric construction with $\phi = 0.7786$ [4], which was subsequently compressed numerically to $\phi = 0.7820$ [5][6]. Here we show that tetrahedra pack much better than this, and in a completely unexpected way. Following a conceptually different approach, using thermodynamic computer simulations that allow the system to evolve naturally towards high-density states, we observe that a fluid of hard tetrahedra undergoes a first-order phase transition to a dodecagonal quasicrystal [7][8][9][10], which can be compressed to a packing fraction of $\phi = 0.8324$. By compressing a crystalline approximant of the quasicrystal, the highest packing fraction we obtain is $\phi = 0.8503$. If quasicrystal formation is suppressed, the system remains disordered, jams, and compresses to $\phi = 0.7858$. Jamming and crystallization are both preceded by an entropy-driven transition from a simple fluid of independent tetrahedra to a complex fluid characterized by tetrahedra arranged in densely packed local motifs that form a percolating network at the transition. The quasicrystal that we report represents the first example of a quasicrystal formed from hard or non-spherical particles. Our results demonstrate that particle shape and entropy can produce highly complex, ordered structures.**

The packing of shapes has drawn the attention of humankind since ancient times. A Sanskrit work from 499 CE reveals the first known mathematical study of the face-centred cubic (FCC) arrangement of spheres [11]. Kepler conjectured and Hales only recently proved the sphere close packing (SCP) fraction of $\phi = \pi/\sqrt{18} \approx 0.7405$ achieved by FCC and its stacking variations [12]. Much less is known about the packing of other shapes. In the case of ellipsoids, periodic arrangements were found with packing fractions up to $\phi = 0.7707$ [3]. With the recent progress in the synthesis of non-spherical particles of sizes ranging from nanometres to microns [13], the problem of packing three-dimensional shapes such as tetrahedra [14] is one of intense interest.

In hard particle systems, the potential energy of two particles is considered infinite if they interpenetrate and zero otherwise. Since all permissible configurations of such systems have the same energy, the equilibrium structure at constant volume maximizes entropy. Surprisingly, hard particles can maximize entropy by ordering. Entropy-driven disorder/order transitions – first predicted by Onsager [15] for hard thin rods and Kirkwood [16] for hard spheres – are now well established for the originally controversial case of spheres, as well as for rods [17], ellipsoids [18], and other shapes [19][20]. In the limit of infinite pressure, an arrangement with maximum packing fraction is stable because it minimizes specific volume and Gibbs free energy.

One of the simplest shapes for which the packing problem is still unsolved is the regular tetrahedron. Tetrahedra do not tile Euclidean space. However, if extra space is allowed between tetrahedra, or between groups of tetrahedra, dense ordered structures become possible. Imagine building a dense cluster, one tetrahedron at a time. As shown in Figure 1(a), a pentagonal dipyramid (PD) is easily built from five tetrahedra if one allows an internal gap of 7.36°. Two PDs can share a single tetrahedron to form a nonamer. Twelve interpenetrating PDs define an icosahedron with a gap of 1.54 steradians. In the figure, tetrahedral dice are stuck together with modelling putty, which distributes the gap that would be present in each motif if most of the adjacent faces were touching. PDs and icosahedra are locally dense, but exhibit non-crystallographic symmetries. The problem of extending or arranging them into space-filling bulk structures is non-trivial. For example, adding a second shell to the icosahedron generates a larger cluster with icosahedral symmetry and 70 tetrahedra, but decreases the packing fraction. For later use, we introduce here a dense, one-dimensional packing given by a linear arrangement of tetrahedra with touching faces known as a tetrahelix, or Bernal spiral.

Recent theoretical works have discussed possible ordered phases of hard tetrahedra formed by some of these motifs. Conway and Torquato [21] proposed the Scottish, Irish, and Welsh configurations, derived from the polytetrahedral networks of clathrate hydrates with packing fractions of up to $\phi = 0.7175$. Chen [4] constructed a crystalline structure formed from nonamers with $\phi = 0.7786$, the first to exceed SCP and showing that tetrahedra obey Ulam's conjecture. Chen's structure was subsequently compressed to $\phi = 0.7820$ [5]. All these packings originated from either geometric considerations or numerical compression. No simulation or experiment has yet reported the spontaneous formation of an ordered phase of hard tetrahedra. Aside from studies of packing, hard tetrahedra have been used to model the structure of water [22]. Polytetrahedral networks of atoms are characteristic for Frank-Kasper phases [23], common in intermetallic compounds.

To obtain dense packings of hard regular tetrahedra, we carry out Monte-Carlo (MC) simulations. Figure 1(b-d) shows the densest configuration ($\phi = 0.8324$) we obtained by first equilibrating an initially disordered fluid of 13824 tetrahedra at constant $\phi = 0.5$ and then compressing the ordered structure that forms. As demonstrated below, this structure is a quasicrystal, with a packing fraction much greater than all previously proposed arrangements of regular tetrahedra. First, we discuss the interesting thermodynamics of the hard tetrahedron fluid.

Figure 2(a) shows the equation of state $\phi(P^*)$ obtained from simulations of a small system with 512 tetrahedra and a large system with 4096 tetrahedra. Here, $P^* = P\sigma^3/k_BT$ is the reduced pressure and $\sigma$ the edge length of a tetrahedron. For the small system, the equilibrium packing fraction exhibits an S-shaped transition at $P^* = 58$ and $\phi = 0.47$ from a simple fluid to a more complex fluid discussed below. At higher pressure the system jams (Figures S1, S2) and, when compressed to nearly infinite pressure, attains a maximum packing fraction of $\phi = 0.7858$. The large system undergoes a first order transition on compression of the fluid phase and forms a quasicrystal. In Figure 2(b), we analyze the system for the presence of locally dense motifs introduced in Figure 1(a). We see that the fraction of tetrahedra belonging to at least one PD increases well before jamming or crystallization. With increasing pressure, interpenetrating PDs form icosahedra and finally merge into a percolating PD network (Fig. 2(c,d)) as the fraction of tetrahedra in PDs approaches unity. For the large system, the fraction of tetrahedra in icosahedra suddenly drops at $P^* = 62$ when crystallization occurs. Comparison with the glass shows that many fewer icosahedra remain in the quasicrystal. Figure 2(c,d) suggests a percolation transition of the PD network in both systems at $P^*_p = 58\pm2$, prior to both jamming and crystallization. We do not observe tetrahedratic liquid crystal phases, which have been suggested by theory [24].

Structural changes of the fluid are revealed by the unusual behaviour of its radial distribution function $g(r)$, as shown in Figure 2(e). We find that the first peak near $r = 0.75\sigma$ disappears upon compression at low pressure, only to reappear for higher pressure, splitting into two peaks at $r = 0.55\sigma$ and $r = 0.80\sigma$. The positions of these peaks are characteristic of face-to-face and edge-to-edge arrangements, respectively, within a single PD. This initial loss of structure with increasing pressure or packing fraction is strikingly different from the well-known behaviour of the hard sphere system depicted in Figure 2(f), and underscores the influence of shape in dense packings.

The spontaneous formation of a quasicrystal from the fluid is remarkable since all previously observed crystalline structures of hard particles have unit cells consisting of only a few particles [17][19][20]. From Figure 1(c) it can be seen that the quasicrystal consists of a periodic stack of corrugated layers with spacing $0.93\sigma$. The view along the direction of the stacking vector (Figure 1(d)) reveals details of the structure within the layers. Twelve-fold symmetric rings formed by interpenetrating tetrahelices exist throughout the structure. The helix chirality is switched by 30° rotations, lowering the symmetry and resulting in a generalized point group of $D_{6d}$ [25].

The structure of the quasicrystal can be understood more easily by examining the dual representation constructed by connecting the centres of mass of neighbouring tetrahedra. In the dual representation, PDs are represented by pentagons. The mapping is applied to a layer of an 8000 particle quasicrystal in Figure 3(a). Recurring motifs are rings of twelve tetrahedra that are stacked periodically to form "logs" (Figure 3(b)), similar to the hexagonal antiprismatic clusters in the tantalum-tellurium system [10]. As indicated in Figure 3(a), the symmetry axes of the logs arrange into a non-repeating pattern of squares and triangles (tile edge length 1.83σ) – an observation that we confirm in systems with 13824 and 21952 particles (Figures S3, S4). The diffraction pattern obtained by positioning scatterers at the centres of tetrahedra shows rings of Bragg peaks, indicating the presence of long-range order with twelve-fold symmetry not compatible with periodicity. Perfect quasicrystals are aperiodic while extending to infinity; they therefore cannot be realized in experiments or simulations, which are, by necessity, finite. The observed tilings and diffraction patterns with twelve-fold symmetry are sufficient in practice for the identification of our self-assembled structures as dodecagonal quasicrystals. Such an identification is in concordance with previous theoretical analysis of random square-triangle tilings [26] and findings of dodecagonal quasicrystals in recent experiments [7][8][9][10] and simulations [27][28].

Quasicrystal approximants are periodic crystals with local tiling structure identical to that in the quasicrystal [25]. Since they are closely related, and they are often observed in experiments, we consider them as candidates for dense packings. The dodecagonal approximant with the smallest unit cell (space group $P\bar{4}n2$) has 82 tetrahedra (Figure 3(c)) and corresponds to one of the Archimedean tilings [29]. At each vertex we see the logs of twelve-member rings (shown in red) capped by single PDs (green). The logs pack well into squares and triangles with additional, intermediary tetrahedra (blue). The vertex configuration of the tiling is $(3,4,3^2,4)$ as shown in Figure 3(d). Interpenetrating tetrahelices can also be seen in the approximant (Figure 3(e)). "Building" and numerically compressing a unit cell of this ideal structure achieves a packing fraction of $\phi = 0.8479$. If we compress a 2x2x2 unit cell, the packing fraction marginally increases to $\phi = 0.8503$, the densest packing of tetrahedra yet reported (Figures S5, S6). Compressing approximants with more complex unit cells, more faithful to an ideal quasicrystal, does not further improve the packing (Figure S7), which suggests that a $(3,4,3^2,4)$ crystal is the thermodynamically preferred phase at our highest pressures.

The fact that the dodecagonal quasicrystal routinely forms in isochoric MC simulations of fluids at packing fractions $\phi \geq 0.5$ indicates that the quasicrystal is thermodynamically favored over the fluid at intermediate pressures. Whether it is stable or metastable relative to the approximant at these pressures is an open question, since the higher tiling entropy of the quasicrystal competes with the higher density of the (3,4,3²,4) approximant (Fig. 2a) to minimize the Gibbs free energy, and entropically stabilized quasicrystals are known to exist [30][31]. Nonetheless, because the transformation to an approximant is a slow process [26], the dodecagonal quasicrystal might be "practically" stable, even if it is not the thermodynamically stable phase.

Why should square-triangle tilings be preferred for dense packings of tetrahedra? First, we compare the packing fraction of the square tile (22 tetrahedra) to that of the triangle tile (9.5 tetrahedra). Their ratio $\phi_{\text{Triangle}} / \phi_{\text{Square}} = 19 / 11\sqrt{3} \approx 0.9972$ is nearly unity, which suggests that tetrahedra pack equally well in both tiles. Second, we note that rings comprising the logs are tilted (Figures 3(b), S8) and the layers of the structure are corrugated (Figures 1(c)). This is a direct consequence of the face-to-face packing of tetrahedra where neighbouring logs kiss. As a result, the square tile has a negative Gaussian curvature while the triangle tile has a positive one. Alternating the two tiles produces a net zero curvature in the layers, as observed in the quasicrystal and its approximant.

As shown in Figure 3(f), the local structure of the (3,4,3²,4) approximant, the dodecagonal quasicrystal and the disordered glass, as characterized by their radial distribution functions are very similar. The peak positions are identical: only the peak heights differ. This implies that the local structure of the glass and quasicrystal are only subtly different, and more sensitive measures of local order, as in Figure 2(b), are required. The crucial step during crystallization is the transformation of the percolating PD network into layers, and the elimination of icosahedra. This intriguing process will be investigated in subsequent studies.

In conclusion, we report the highest known packing fraction of regular tetrahedra and show unexpected ways in which they can pack more densely than previously proposed, including the first quasicrystal formed from non-spherical particles. The spontaneous formation of a quasicrystal of hard particles demonstrates that shape alone can produce remarkable structural complexity through solely entropic interactions.

*Note added in proof:* We are pleased to note the as-yet-unpublished posting of a new result by Kallus *et al.* [32] on the dense packing of tetrahedra in a dimer crystal, indicating a packing fraction of $100 / 117 \approx 0.8547$.

## METHODS SUMMARY

We use isobaric and isochoric MC simulations with periodic boundary conditions to study systems of $N$ regular tetrahedra, with $N$ ranging from 512 to 21952. A full MC cycle consists of $N + 1$ trial moves including translation plus rotation of a tetrahedron or rescaling of the orthorhombic box. Maximum step sizes are updated occasionally to keep the acceptance probabilities at 30%. Simulations are initialized at low packing fraction in a random configuration and subsequently squeezed to higher densities. The dodecagonal quasicrystals shown in Figures 1, 3, S1, and S2 are obtained in isochoric simulations at packing fraction $\phi = 0.50$. Crystallization proceeds in three steps: (i) equilibration of the dense, metastable fluid (e.g. $N = 8000$: $<12\times10^6$ MC cycles); (ii) nucleation and growth ($12-23\times10^6$ MC cycles); and (iii) healing of defects ($>23\times10^6$ MC cycles). The equation of state in Figure 2(a) is calculated by increasing (decreasing) the external pressure step-wise for compression (decompression). Longer simulations facilitate equilibration in the transition region. For detecting PDs and icosahedra in Figure 2(b), nearest neighbours are sampled with a distance cut-off of $0.65\sigma$. The resulting motifs are further screened by projecting the directions of the tetrahedra onto the surface of the unit sphere, and indexing the resulting pattern using spherical harmonics and comparison with an ideal pattern (pentagon for a PD and dodecahedron for an icosahedron). For $P^* > 120$, compression with conventional MC is inefficient. Therefore we apply an alternative method to reach pressures as large as $P^* \geq 10^6$ and obtain maximum density packings. The method relies on allowing a small number (on the order of 0.1% of all particles) of minor overlaps (interpenetration of tetrahedra) during box rescaling. All overlaps are subsequently eliminated with isochoric MC. Details of our algorithms and the particle data are given in below.

**REFERENCES**

[1] Gardner, M. *The Colossal Book of Mathematics: Classic puzzles, paradoxes, and problems*, p. 135 (Norton, New York, 2001).

[2] Betke, U. & Henk, M. Densest lattice packings of 3-polytopes, *Comput. Geom.* **16**, 157-186 (2000).

[3] Donev, A., Stillinger, F. H., Chaikin, P. M. & Torquato, S. Unusually dense crystal packing of ellipsoids, *Phys. Rev. Lett.* **92**, 255506 (2004).

[4] Chen, E. R. A Dense packing of regular tetrahedra, *Discrete Comput. Geom.* **40,** 214-240 (2008).

[5] Torquato S. & Jiao Y. Dense packings of the Platonic and Archimedean solids, *Nature* **460**, 876-879 (2009).

[6] Torquato S. & Jiao Y. Dense packings of polyhedra: Platonic and Archimedean solids, *Phys. Rev. E* **80**, 041104 (2009).

[7] Zeng, X., Ungar, G., Liu, Y., Percec, V., Dulcey, A. E. & Hobbs, J. K. Supramolecular dendritic liquid quasicrystals, *Nature* **428**, 157-160 (2004).

[8] Hayashida, K., Dotera, T., Takano, A. & Matsushita, Y. Polymeric quasicrystal: mesoscopic quasicrystalline tiling in ABC star polymers, *Phys. Rev. Lett.*, **98**, 195502 (2007).

[9] Talapin, D. V., Shevchenko, E. V., Bodnarchuk, M. I., Ye, X., Chen J. & Murray, C. B. Quasicrystalline order in self-assembled binary nanoparticle superlattices, *Nature* **461**, 964-967 (2009).

[10] Conrad, M., Krumeich, F. & Harbrecht, B. A dodecagonal quasicrystalline chalcogenide, *Angew. Chem. Int. Ed.* **37**, 1384-1386 (1998).

[11] Hales, T. C. Historical overview of the Kepler conjecture, *Discrete Comput. Geom.* **36**, 5-20 (2006).

[12] Hales, T. C. A proof of the Kepler conjecture, *Ann. Math.* **162,** 1065-1185 (2005).

[13] Glotzer, S. C. & Solomon, M. J. Anisotropy of building blocks and their assembly into complex structures, *Nature Mat.* **6,** 567-572 (2007).

[14] Tang, Z. Y., Zhang, Z. L., Wang, Y., Glotzer, S. C. & Kotov, N. A. Spontaneous self-assembly of CdTe nanocrystals into free-floating sheets, *Science* **314**, 274-278 (2006).

[15] Onsager, L. The effect of shape on the interaction of colloidal particles, *Ann. NY Acad. Sci.* **51**, 627–659 (1949).

[16] Kirkwood, J. E. in *Phase transformations in solids* ed. by R. Smoluchowski, J. E. Mayer & W. A. Weyl, p.67 (Wiley, New York, 1951).

[17] Bolhuis, P. & Frenkel, D. J. Tracing the phase boundaries of hard spherocylinders, *J. Chem. Phys.* **106**, 666-687 (1997).

[18] Camp, P. J. & Allen, M. P. Phase diagram of the hard biaxial ellipsoid fluid, *J. Chem. Phys.* **106**, 6681-6688 (1997).

[19] Veerman, J. A. C. & Frenkel, D. Phase-behavior of disk-like hard-core mesogens, *Phys. Rev. A* **45**, 5632-5648 (1992).

[20] John, B. S., Juhlin, C., & Escobedo, F. A. Phase behavior in colloidal hard perfect tetragonal parallelepipeds, *J. Chem. Phys.* **128**, 044909 (2009).

[21] Conway, J. H. & Torquato, S. Packing, tiling and covering with tetrahedra, *Proc. Natl. Acad. Sci.* **103**, 10612-10617 (2006).

[22] Kolafa, J. & Nezbeda, I. The hard tetrahedron fluid: a model for the structure of water, *Mol. Phys.* **84**, 421-434 (1994).

[23] Frank, F. C. & Kasper, J. S. Complex alloy structures regarded as sphere packings. 1. Definitions and basic principles, *Acta Cryst.* **11**, 184-190 (1958).

[24] Fel, L. G. Tetrahedral symmetry in nematic liquid crystals, *Phys. Rev. E* **52**, 702-717 (1995).

[25] Yamamoto, A. Crystallography of quasiperiodic crystals, *Acta Cryst.* **A52**, 509-560 (1996).

[26] Oxborrow, M. & Henley, C. L. Random square-triangle tilings – a model for twelvefold-symmetrical quasi-crystals, *Phys. Rev. B* **48**, 6966-6998 (1993).

[27] Roth J. & Denton A. R., Solid-phase structures of the Dzugutov pair potential, *Phys. Rev. E* **61**, 6845-6857 (2000).

[28] Keys, A. S. & Glotzer, S. C. How do quasicrystals grow? *Phys. Rev. Lett.* **99**, 235503 (2007).

[29] Mikhael, J., Roth, J., Helden, L. & Bechinger, C. Archimedean-like tiling on decagonal quasicrystalline surfaces, *Nature* **454**, 501-504 (2008).

[30] Steurer W. Structural phase transitions from and to the quasicrystalline state, *Acta Cryst.* A**61**, 28-38 (2005).

[31] Engel, M. & Trebin H.-R. Self-assembly of complex crystals and quasicrystals with a double-well interaction potential, *Phys. Rev. Lett.* **98**, 225505 (2007).

[32] Kallus, Y, Elser, V. & Gravel S. A dense periodic packing of tetrahedra with a small repeating unit, arXiv:0910.5226 (2009).

**Supplementary Information** is linked to the online version of the paper at www.nature.com/nature.

**Acknowledgements** We acknowledge support from the Air Force Office of Scientific Research and the National Science Foundation. M.E. was supported by a postdoctoral fellowship of the Deutsche Forschungsgemeinschaft.

**Author Contributions** A.H.-A. and M.E. performed all simulations and contributed equally to the study. M.E. solved the quasicrystal and approximant structures. A.S.K. performed shape-matching analysis. X.Z., P.P.-M, and R.G.P. proposed and constructed geometric packings. All authors discussed and analysed the results, and contributed to the scientific process. S.C.G., A.H-A, and M.E. wrote most of the paper, and all authors contributed to refinement of the manuscript. S.C.G. and P.P.-M. conceived and designed the study, and S.C.G. directed the study.

**Author Information** Reprints and permissions information is available at www.nature.com/reprints. Correspondence and requests for materials should be addressed to S.C.G. (sglotzer@umich.edu).

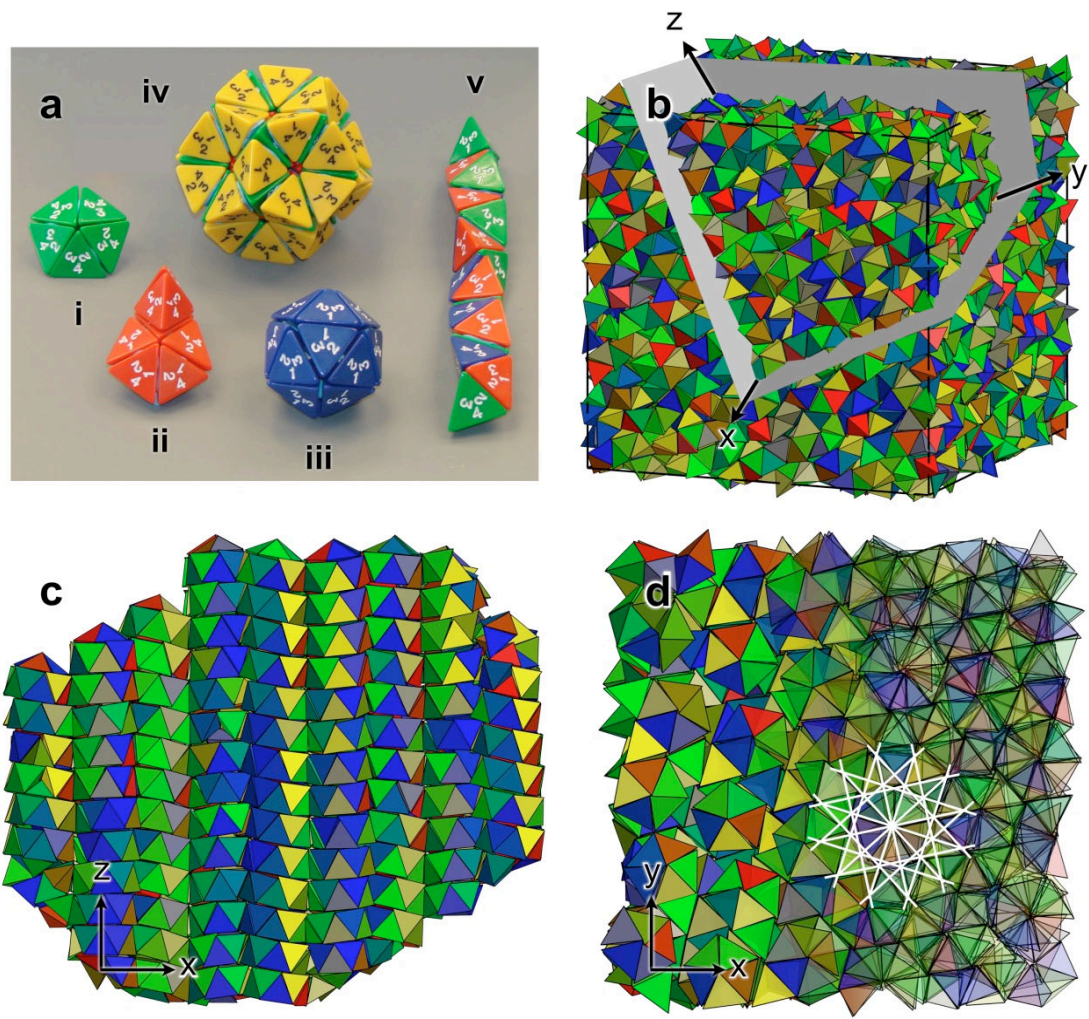

**Figure 1. Packings of tetrahedra obtained by hand and by computer simulation.** (a) Ideal local packing motifs built from tetrahedral dice stuck together with modelling putty: (i) pentagonal dipyramid, (ii) nonamer, and (iii) icosahedron maximize local packing density. The icosahedron can be extended by adding a second shell (iv), but then the large gaps between the outer tetrahedra lower the density. (v) The tetrahelix maximizes packing density in one dimension. (b-d) A quasicrystal with packing fraction $\phi = 0.8324$ obtained by first equilibrating an initially disordered fluid of 13824 hard tetrahedra using Monte Carlo simulation and subsequent numerical compression. The images show an opaque view of the system (b) and opaque and translucent views of two rotated narrow slices (c)-(d). The white overlay in (d) shows the distinctive twelve-fold symmetry of the dodecagonal quasicrystal. Corrugated layers with normals along **z** are apparent in (c). The colouring of tetrahedra is based on orientation.

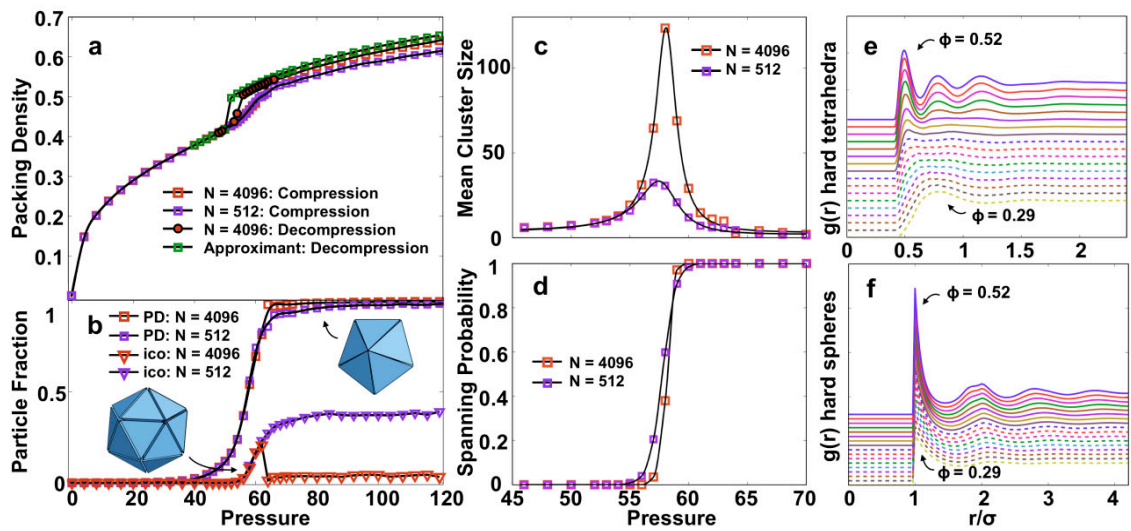

**Figure 2. Thermodynamic and structural properties of the hard tetrahedron fluid.** (a) Equation of state spanning the transition from the liquid to the solid state. Data is shown for various system sizes. Decompression of both the quasicrystal ($N = 4096$) and the approximant (unit cell $N = 82$) shows a sharp melting transition. Hysteresis of the compression and decompression curves for the quasicrystal further indicates a first order transition. For the system with $N = 512$, crystallization is inhibited in many runs, producing a jammed, disordered glass. (b) Fraction of tetrahedra participating in pentagonal dipyramids ('PD', right inset) and icosahedra ('Ico', left inset). (c) Mean cluster size of interpenetrating PDs. (d) Spanning probability of the largest cluster of interpenetrating PDs. (e) Radial distribution function g(r) of regular tetrahedra for packing fractions ranging from $\phi = 0.29$ to $\phi = 0.52$. Curves are vertically offset for clarity. (f) Radial distribution function for the same densities as in (e) for a hard sphere system.

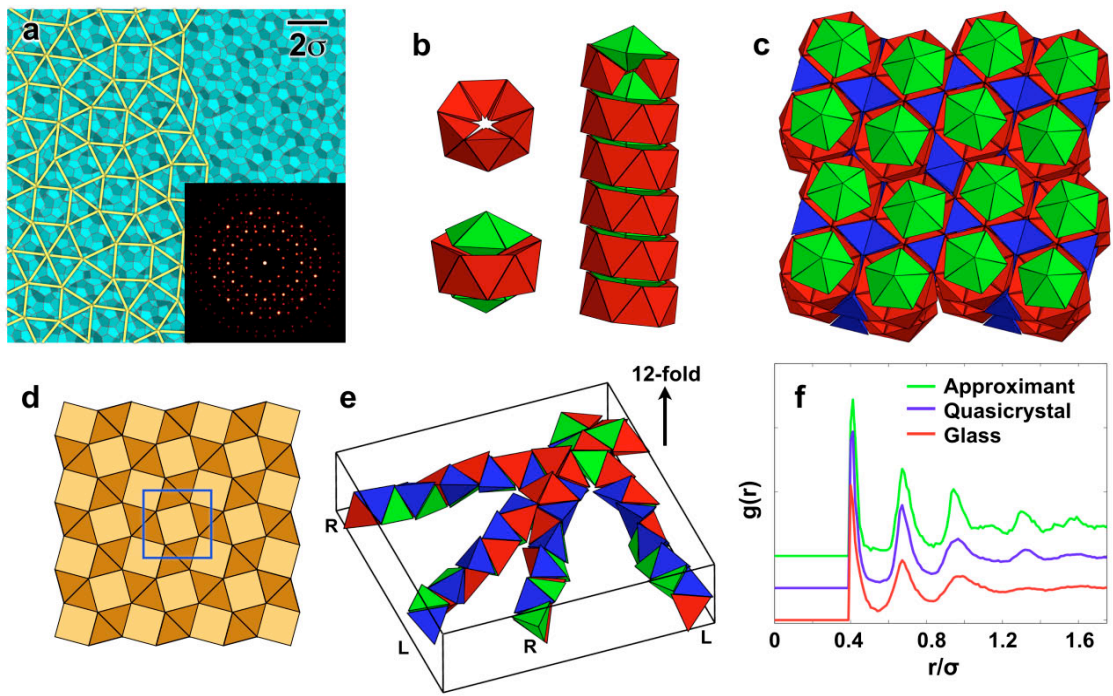

**Figure 3. Structural characterization of the hard tetrahedra dodecagonal quasicrystal and its approximant.** (a) Network of connected tetrahedra in a slice through a quasicrystal with 8000 tetrahedra, viewed along the direction of the twelve-fold axis. Lines connecting the centre of mass of nearest neighbour tetrahedra form turquoise pentagons, which correspond to the PD network. The tiling structure is highlighted in yellow. A diffraction pattern indicating twelve-fold symmetry is shown in the inset. (b) The vertices of the tiling are formed by logs comprised of rings of twelve tetrahedra, with neighbouring rings enclosing a PD. The packing fraction within the logs can be increased by a tilt of the rings with respect to the log axis. This allows neighbouring PDs to avoid each other, as indicated in the figure where two tetrahedra have been removed from the top ring to expose PDs. Structure (c) and tiling (d) of the $(3,4,3^2,4)$ approximant to the dodecagonal quasicrystal of tetrahedra. Colours are described in text. (e) Similar to the quasicrystal, interpenetrating tetrahelices are present throughout the approximant. Their chirality alternates between left (L) and right (R) by 30° rotations. (f) Radial distribution functions for the approximant ($N = 82$), quasicrystal ($N = 8000$), and glass ($N = 8000$). Curves are vertically offset for clarity.

## METHODS

We define an ideal, regular tetrahedron as the convex hull of its four vertices $v_1 = (1, 1, 1)$, $v_2 = (1, -1, -1)$, $v_3 = (-1, 1, -1)$, $v_4 = (-1, -1, 1)$. The edge length of the tetrahedron is $\sigma = \sqrt{8}$ and its volume $V_T = 8/3$. The position and orientation of an arbitrary tetrahedron is given by $(x, y, z, a, b, c, d)$, where $(x, y, z)$ is the translation vector and $(a, b, c, d)$ a quaternion describing the rotation. The relation between the quaternion and the rotation matrix is:

$$R = \begin{pmatrix} a^2 + b^2 - c^2 - d^2 & 2(bc - ad) & 2(bd + ac) \\ 2(bc + ad) & a^2 - b^2 + c^2 - d^2 & 2(cd - ab) \\ 2(bd - ac) & 2(cd + ab) & a^2 - b^2 - c^2 + d^2 \end{pmatrix}.$$

**Overlap detection**

Individual MC moves consist of small translation steps and small rotation steps within a cubic box of side length $\Delta r$. The most time-consuming part of our algorithm is the overlap check. Tetrahedra are sorted into a cell list with cell size $2\sqrt{3}$, which is the distance beyond which two tetrahedra cannot overlap. Two different overlap detection algorithms are used. They have been designed, written, and tested independently by different co-authors of the present work. Comparing the results of the algorithms on a test set with overlapping and non-overlapping tetrahedra allows independent verification of the codes. The numerical precision of the overlaps detection algorithm and therefore the reported packing densities is standard double floating-point precision.

The first algorithm is based on the observation that two convex polyhedra overlap, if and only if at least one edge of one polyhedron intersects one face of the other polyhedron. This means that maximally $2 \times 6 \times 4 = 48$ intersections of lines and triangles have to be evaluated. The algorithm can be optimized by sorting the vertices of each tetrahedron based on their distance from the centre of the other tetrahedron. Only the three edges among the three closest vertices for each tetrahedron require inspection. Additionally, the face that does not include the nearest of the vertices can be discarded. This reduces the necessary checks to a maximum of 18.

The second, independent algorithm takes advantage of the fact that two convex polyhedra do not intersect if and only if a plane can be found that completely separates them. In other words, the vertices of one tetrahedron must lie on one side of the plane and the vertices of the other tetrahedron on the other side. Taking two vertices of one tetrahedron and one vertex of the other tetrahedron defines a trial candidate for a separating plane. It can be shown that the study of all such trial candidates is sufficient. Thus, 6 × 2 × 2 = 24 trial candidates need to be checked. Similar to the first algorithm, sorting the vertices allows reduction in the number of candidates. For production simulation runs we use the second overlap detection algorithm only because it is more efficient than the first algorithm.

**Compression algorithm**

To obtain the maximum density for a given configuration, we quickly compress to very high densities a system first equilibrated using standard isobaric or isochoric MC. Since such rapid compression is inefficient with the standard isobaric MC scheme, we use a modified MC scheme. Our modified scheme is only used to obtain high-density results; the conventional isochoric and isobaric MC methods are used to produce equation of state data and to produce the quasicrystals and jammed structures from the fluid.

In the conventional isobaric MC algorithm for hard particles, trial volume changes are performed by rescaling the box dimensions. If such a volume change creates an overlap it is discarded, otherwise it is accepted according to the Metropolis criterion for the isobaric ensemble. As the density increases, trial compression moves generate overlaps with larger and larger probability, especially for big systems, making conventional isobaric MC slow in equilibrating high densities. To compress our system more efficiently, we introduce a modified scheme that allows a few minor overlaps during compression; these overlaps are then removed to obtain the final configuration.

For our modified scheme, we always accept volume changes, even if they create overlaps. To ensure that the number and amount of overlaps remains small, we use a separate criterion to decide whether the box should be expanded or compressed. We keep track of the acceptance probability $p$ of MC translation moves and compare this to a target acceptance probability $p_0 = 0.3$. If $p < p_0$, we apply a compression move, otherwise we apply an expansion move. For our system, we find that rescaling of the box dimensions by a random factor in between 1 and 1 + 0.002 $\Delta r$ for expansion and a random factor between 1 − 0.002 $\Delta r$ and 1 for compression gives a good balance between fast compression and creating overlaps that are small enough to remove later. The average fraction of particles that overlap during compression is very small, on the order of 0.1% of all particles.

The control parameter for our "high-density" compression algorithm is the maximum distance of MC translation moves, $\Delta r$. Although it is not possible to directly measure or control the pressure in our method, we observe that a lower value of $\Delta r$ corresponds to a higher pressure. For maximum compression, $\Delta r$ is lowered exponentially to zero.

Our method is extremely simple, fast, and robust. By running a short isochoric simulation at the end, we instantaneously remove any pre-existing overlaps. This involves implementing the standard translation and rotation MC moves, which guarantees that no new overlaps are created and allows sufficient motion, even at very high densities. Using the two independent overlap detection algorithms described above, we ensure that all overlaps are removed for the data reported in this Letter.

**Performance**

On a single CPU core, the run time for our most efficient overlap detection scheme is about 5 μs per particle per MC cycle at an intermediate packing density. Typical compression runs for the 13824 tetrahedra system take on the order of a few hours on a single AMD Opteron CPU core with 2.3 GHz clock speed. To obtain good candidates with maximum densities we extended the compression time to a few days. Due to these finite compression times, we restrict the packing densities given in the text to four significant digits. Nucleating the quasicrystal from the fluid in an isochoric simulation and then compressing the quasicrystal with 13824 particles over 40 million MC cycles using our compression algorithm to achieve the data shown in Figure 1 (b-d) took about one month on a single CPU core.

**SUPPLMENTARY INFORMATION**

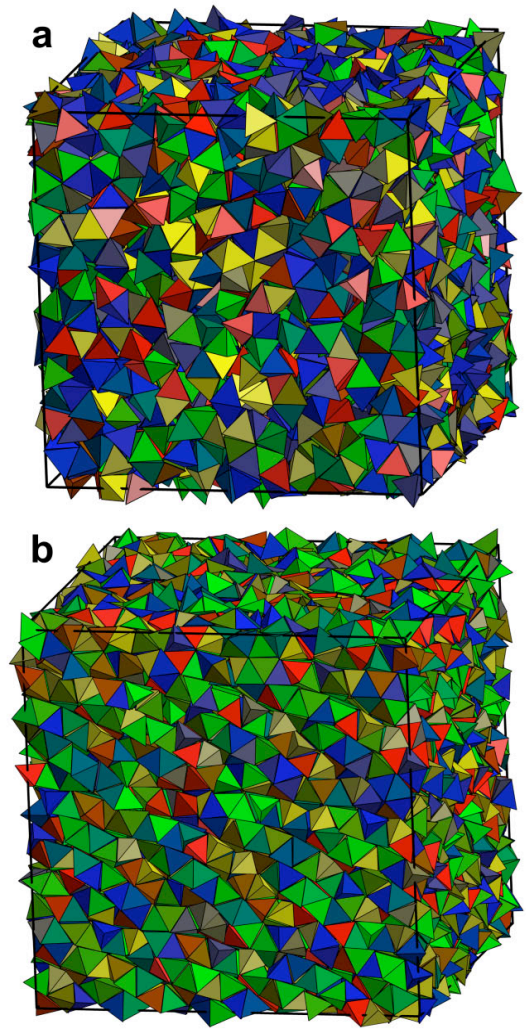

**Figure S1. Snapshots of large systems after the final compression to very high pressures.** (a) In short simulations or if the box is compressed too rapidly, crystallization does not occur. The image shows a disordered arrangement of $N = 8000$ tetrahedra, compressed to $\phi = 0.7858$. Local ordering of the tetrahedra is visible. (b) A quasicrystal with $N = 13824$ was grown in an isochoric simulation at $\phi = 0.5$. In the subsequent compression the density increased to $\phi = 0.8324$. The periodic layers of the dodecagonal quasicrystal run from the bottom right to the top left. The twelve-fold axis points to the top right. The bent sequences of tetrahedra visible in the image are an artifact of the straight cut by the simulation box. These are not the periodic layers.

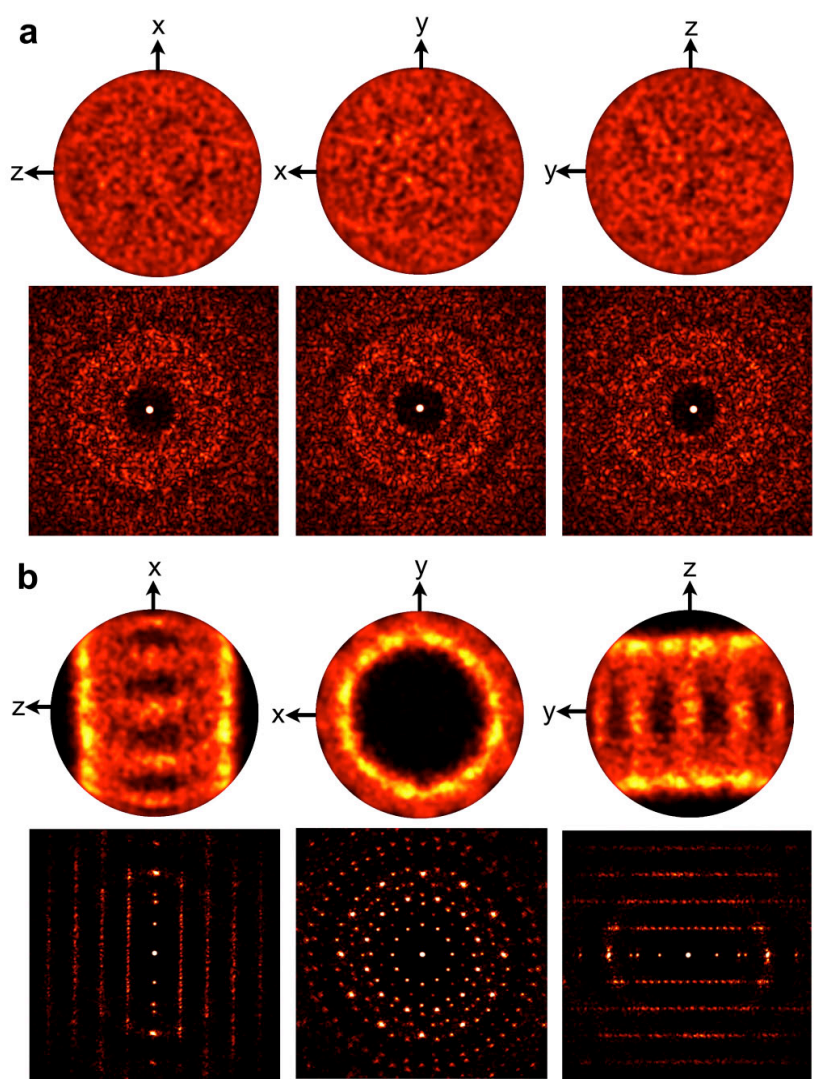

**Figure S2**. **Comparison of the global order in the dense glass with the global order in the dense quasicrystal.** The top rows show bond order diagrams, which are obtained by projecting the vectors connecting the centres of nearest neighbour tetrahedra separated by ≤ 0.55σ on the surface of a sphere. The bottom rows depict diffraction patterns. (a) For the glass (Figure S1(a)), bond directions are distributed uniformly on the sphere, and the diffraction patterns are isotropic. (b) Due to the layering of the quasicrystal (Figure S1(b)), very few bonds are observed to form angles of ≤ 60° with the z-axis. Bonds with small angles relative to the x-y-plane belong to tetrahelices. Peaks in the diffraction pattern are confined to equidistant planes perpendicular to the twelve-fold axis, verifying that the system is periodic in this direction.

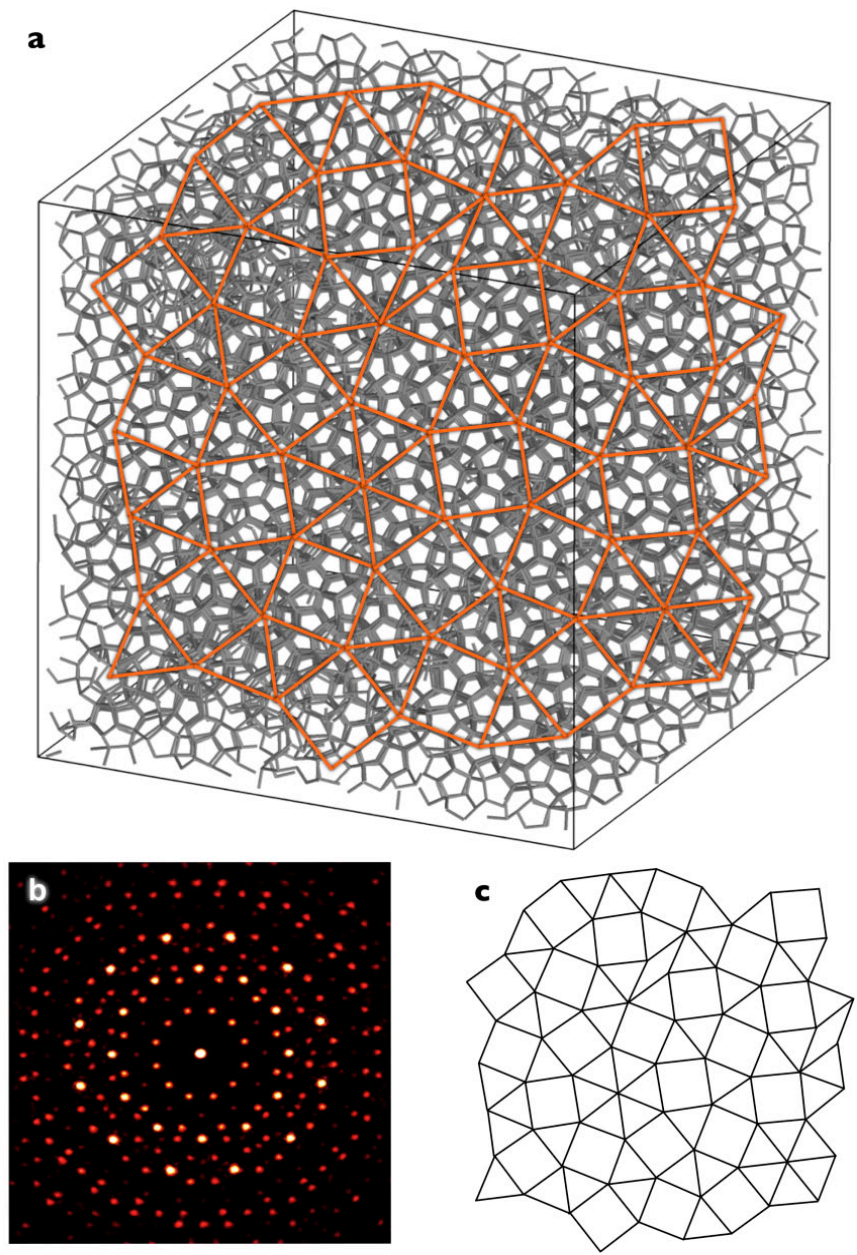

**Figure S3. Analysis of the tiling substructure in the quasicrystal with $N = 24^3 = 13824$ particles shown in Figure 1(b-d).** (a) The gray lines connect tetrahedra that are nearest neighbours according to the first peak in the radial distribution function. Pentagons in the neighbour network correspond to PDs. Centres of logs are connected with orange lines. (b) The diffraction pattern shows twelve-fold symmetry. (c) The square-triangle tiling contains $n_S = 23$ squares, $n_T = 56$ triangles and $n_R = 2$ thin rhombi. Their ratio $n_T / (n_S + n_R / 2) \approx 2.33$ is close to the ideal value for a mathematically perfect quasicrystal, $4 / \sqrt{3} \approx 2.31$. For the approximant the ratio equals 2.

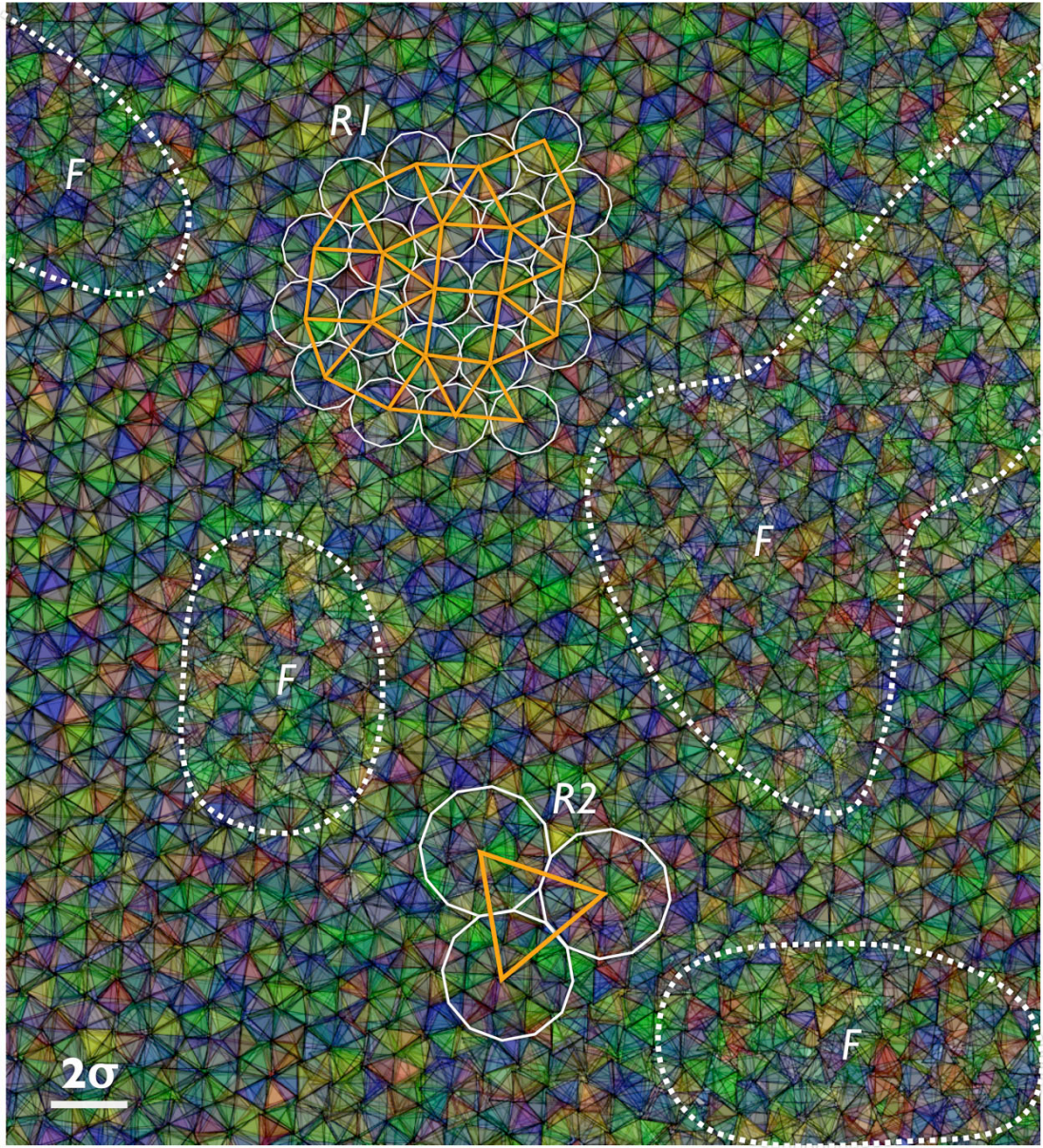

**Figure S4. Translucent view showing a cut through a system with $N = 28^3 = 21952$ tetrahedra.** The viewing direction is along the dodecagonal axis. Parts of the sample are still in the fluid (*F*) state. Tetrahedra arrange into small rings (*R*1) and large rings (*R*2). Their diameters have an irrational ratio of $r_2 / r_1 = (\sqrt{3} + 1) / \sqrt{2}$. The twelve-fold logs discussed in the main text correspond to *R*1. As indicated with orange lines, the rings form squares and triangles, which are arranged into a random dodecagonal square-triangle tiling.

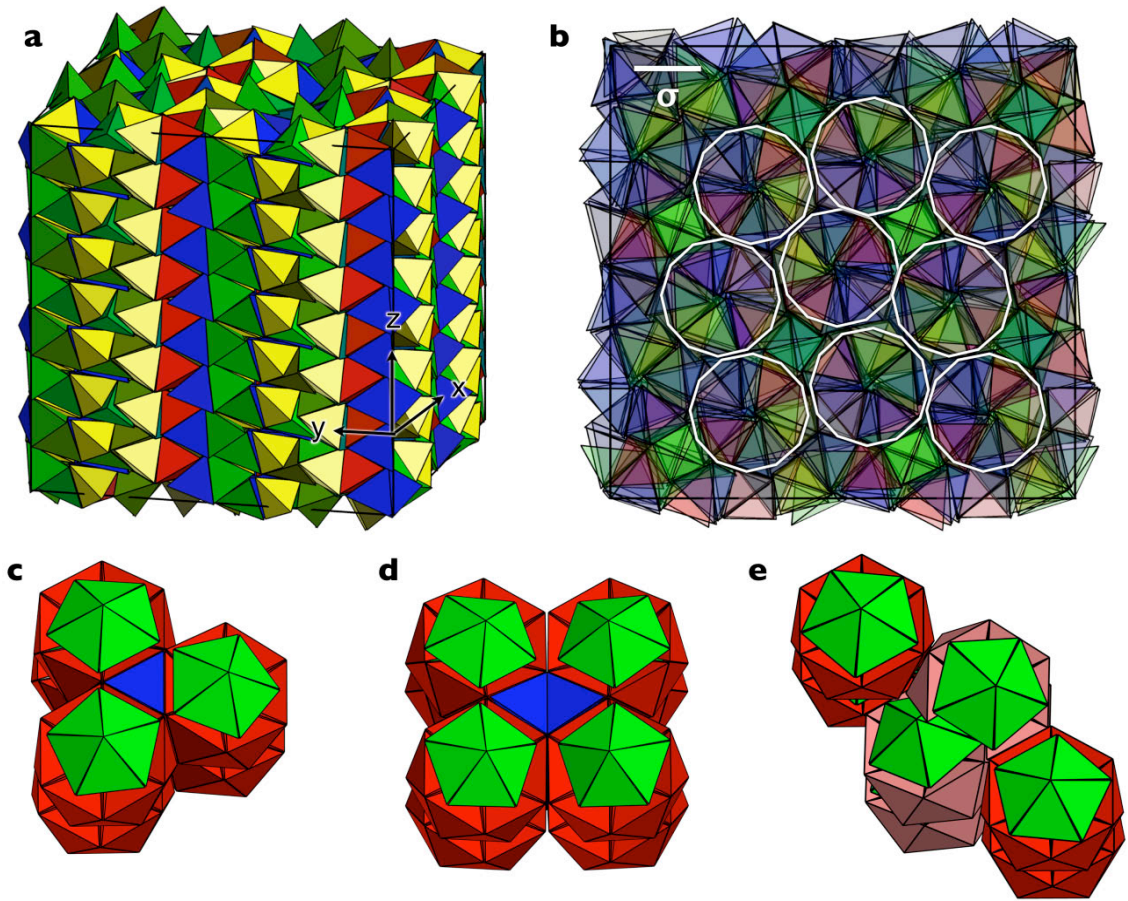

**Figure S5**. **Details of the highest density ($\phi$ = 0.8503) packing of hard tetrahedra observed in this study.** This density was obtained by compressing a 2×2×2 cell of the (3,4,3$^2$,4) approximant with 656 tetrahedra. (a) For ease of viewing, the 2×2×2 cell has been periodically continued into a 2×2×8 cell. (b) In the translucent image, the twelve-fold logs can be identified. The logs arrange into triangle (c) and square (d) tiles with 9.5 and 22 tetrahedra, respectively. The ratio of the packing densities of the tiles is $\phi_{\text{Triangle}} / \phi_{\text{Square}} = 19 / 11\sqrt{3}$. (e) Thin rhombi are frequently observed in connection with zipper motion, a dynamical mechanism to rearrange squares and triangles [26]. A rhomb is a structural defect. It consists of 11 tetrahedra and has half the volume of a square. As can be seen in the image, tetrahedra in the middle form a spiral of two interpenetrating logs (light red).

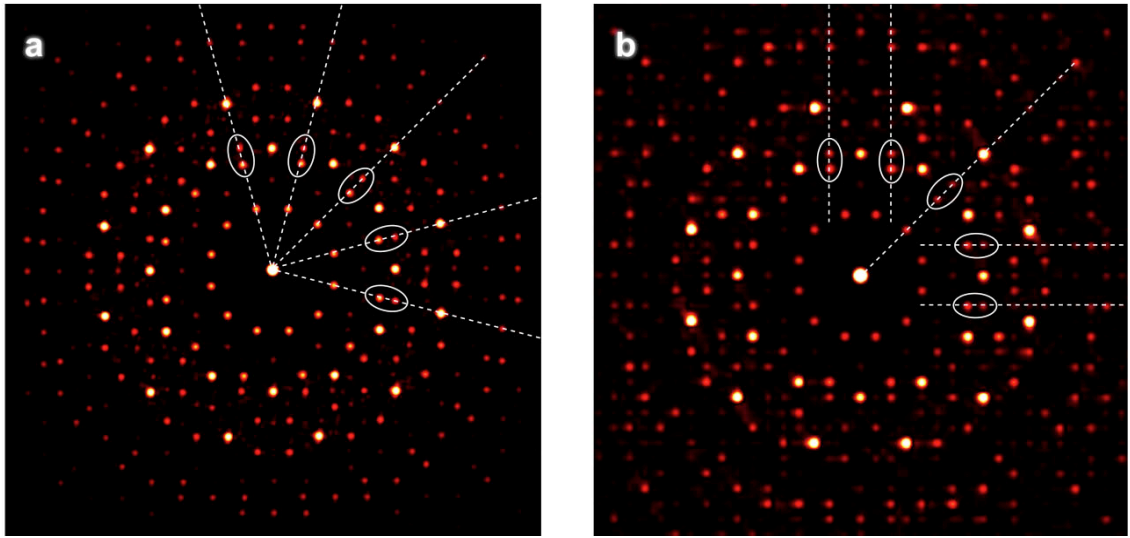

**Figure S6. Comparison of the diffraction patterns of the compressed quasicrystal shown in Figure 1 and the compressed (3,4,3$^2$,4) approximant shown in Figure S5.** While the Bragg peaks have perfect twelve-fold symmetry in the dodecagonal quasicrystal (a), the symmetry is broken to four-fold symmetry in the approximant (b). As indicated by white dashed lines and ellipsoids, weak Bragg peaks of the approximant are shifted slightly from their positions in the quasicrystal. Such behavior can be understood within the theory of quasicrystals [25].

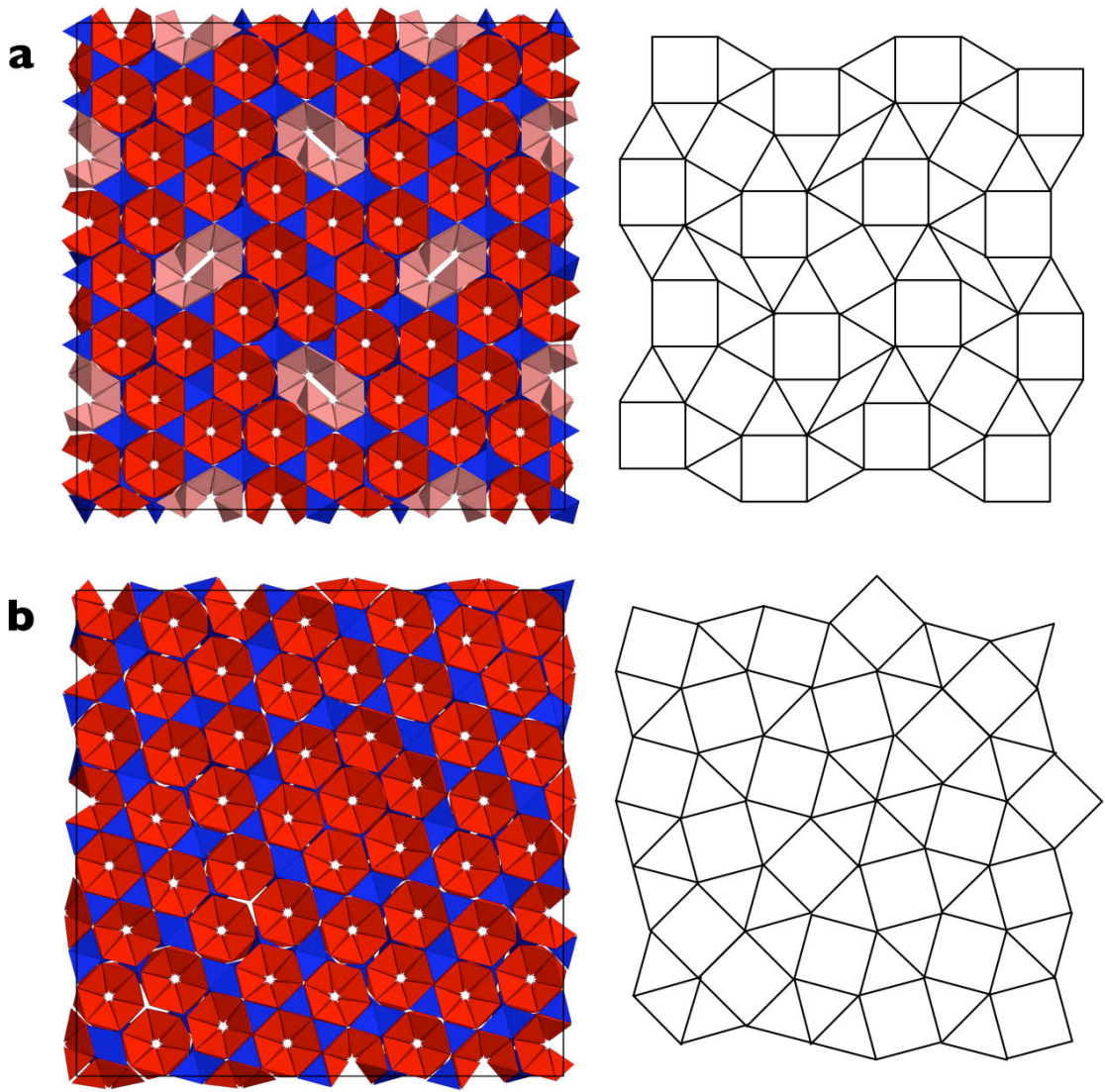

**Figure S7. Two higher approximants of the dodecagonal quasicrystal obtained by geometric construction.** Arrangements of tetrahedra (left) and tilings (right) are shown. PDs capping the logs are not drawn for ease of viewing. (a) The second approximant has a body-centred tetragonal unit cell with 306 tetrahedra. Four unit cells form an orthorhombic box. (b) The third approximant has a primitive tetragonal unit cell with 1142 tetrahedra. As expected, the ratio of the number of tetrahedra in successive approximants converges to the self-similarity scaling factor of the square-triangle tiling, $\sqrt{3} + 2 \approx 3.732051$: $306 / 82 \approx 3.731707$ (2nd vs. 1st), $1142 / 306 \approx 3.732026$ (3rd vs. 2nd). The approximants compress to $\phi = 0.8284$ (a) and $\phi = 0.8352$ (b).

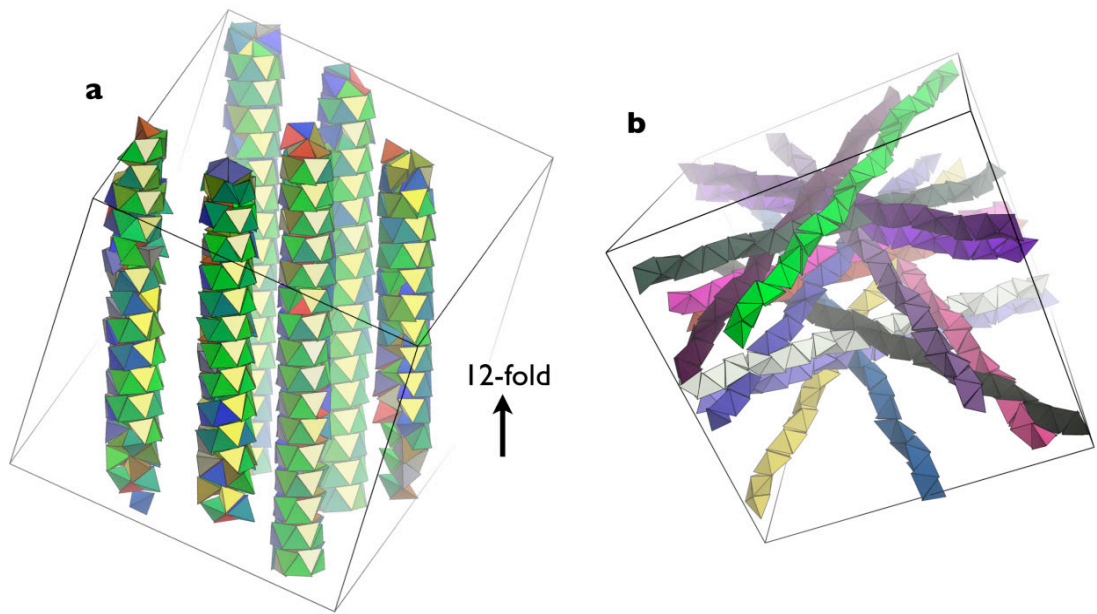

**Figure S8**. **One-dimensional building blocks of the dodecagonal quasicrystal identified in the $N = 13824$ system (Figure 1(b-d)).** (a) Tetrahedra forming six randomly selected logs are extracted from the sample. Periodic ordering along the twelve-fold axis and slight tiltings of the rings with respect to the log axis are visible. (b) Tetrahelices containing more than 48 tetrahedra (not taking into account periodic boundary conditions) are displayed. When projected along the twelve-fold axis, the orientations of the tetrahelices are limited to twelve directions. Both left-handed and right-handed tetrahelices can be seen. The chirality is observed to switch every 30 degrees. This can be understood from the geometry of the rings within logs: A 30 degree rotation around the log axis is the same as a horizontal reflection. Since a reflection switches chirality, the chirality of the tetrahelix is switched every 30 degrees.

**Tetrahedra coordinates and orientations**

We list the positions of the tetrahedra centres (rows 1 to 3) and orientations (rows 4 to 7) of the 82 particle unit cell (3,4,3$^2$,4) approximant with packing density $\phi = 0.8478659094$ for copy and paste. Orientations are given in the form of quaternions as described in the Method Section of the manuscript. In the first two lines the box dimensions and the numbers of particles are given. The compression run covered 20 million MC simulation cycles.

```
10.2088453514   2.5716864401    9.8233760813
82
0.5383279474   −1.0816365671   −0.0346011490    0.8173375866    0.3278806957    0.2967923896    0.3692801056
−1.0385101649  −1.1137606006   −1.2578328424    0.5724095042    0.5550696822    0.5868105193    0.1410617658
−2.9362480682  −0.7609530428   −0.8577187102    0.1812308465    0.6275717323    0.7442250121   −0.1394210617
−3.1890605156  −0.6103678331    1.0837087544   −0.2628084292    0.5555971254    0.6966422770   −0.3700447293
−1.7007928845  −0.5960376242    2.3434824230   −0.6060764129    0.3134568538    0.5104765500   −0.5232875638
 0.0933213509  −0.7784939675    1.7805610499   −0.8318710875   −0.0026680948    0.0970341615   −0.5464135307
−2.7679381132  −1.2323019866    1.9825306380   −0.5114667801    0.1108778509    0.1677973563    0.8354351455
−0.7613453994   1.2565044539    2.3933605449   −0.4870445376    0.4823672469   −0.1420120058    0.7141022670
 0.5494142589   0.9925532575    1.0846369909   −0.3033868559    0.7461410488   −0.4572368682    0.3770469431
−0.1251523930   0.8237230804   −0.7751415814   −0.0568358041    0.7527695353   −0.6556961775   −0.0130476419
−2.1946421997   0.9879520386   −1.2260875466    0.1595793590    0.6234804934   −0.6425373272   −0.4158753247
−3.3812438776   1.2297675815    0.0738936542    0.3931249423    0.2553417994   −0.4531678517   −0.7582164885
−0.7081137817  −0.0214954395   −0.1968296922    0.8779254263    0.3076699780    0.3002025087   −0.2108662712
−1.8579072647   0.0413044774   −0.3565562765    0.5062046296    0.2157859331    0.7319016593   −0.4018871298
−2.4355724289   0.2336515817    0.6820981696   −0.0772856769    0.0276858029    0.8907704315   −0.4469770227
−1.4775507487   0.3397319832    1.4199362142   −0.6321643113   −0.1653752875    0.6896399490   −0.3121154251
−0.4525815529   0.1547955816    0.9213487609   −0.9251261494   −0.2891487777    0.2374926589   −0.0642793052
−4.6960666301  −0.8844649422   −1.3463660315    0.7527987955    0.2947406577    0.3424764514    0.4786771338
 4.1348734545  −1.1755657333   −2.5585872154    0.5107492370    0.4847866980    0.6878588346    0.1759752765
 2.0317644767  −1.2433195246   −2.1377909308    0.2043144333    0.5158456725    0.8245817642   −0.1105611533
 1.6585076219  −1.1377233428   −0.3134037530   −0.2374141510    0.3955545034    0.7683447450   −0.4436411936
 3.2715417031  −0.8760383382    0.9659086251   −0.5419996312    0.2506555680    0.4945770155   −0.6315075310
−4.9891652919  −0.5936596571    0.5178340545   −0.7566644218   −0.0197442114    0.1018465153   −0.6455202602
 2.3603607416   0.9564777469    0.5132605556   −0.6429538838    0.1361197087    0.1568831921    0.7372037657
 4.3771279136   1.2850478525    1.0823639250   −0.5622847378    0.5280526662   −0.0719906965    0.6323081488
−4.5220078305   1.1772645849   −0.2274739518   −0.3504398541    0.8047585521   −0.3561058563    0.3205373623
 5.0428109541   0.8788146073   −2.0520474288   −0.0429462841    0.8367547787   −0.5426721052   −0.0591949597
 2.9677362922   0.7217344912   −2.5231741276    0.2153804229    0.7076811573   −0.5483934537   −0.3899529112
 1.7069199696   0.6347178717   −1.1957036369    0.4952202298    0.3394430087   −0.4178464548   −0.6818648752
 4.4581478959   0.0380822377   −1.3601903231    0.9097007557    0.2714648157    0.3084844592   −0.0599059879
 3.3779478506  −0.1904382036   −1.6991501637    0.5891510945    0.1247841751    0.7419994725   −0.2945618447
 2.6798376682  −0.2107133522   −0.7731601623    0.0500655532   −0.0704763497    0.9021882619   −0.4225906585
 3.2996417100  −0.0162351737    0.1888022857   −0.5027974922   −0.2545024717    0.7262921663   −0.3936024173
 4.4346061708   0.2090920145   −0.1827233633   −0.8861390195   −0.3227680700    0.2490381693   −0.2203597089
 1.8387537496  −1.1463583386   −4.7995371856    0.7615454298    0.4389956286    0.2031533858    0.4313468422
 0.5655152773  −0.8828117021    3.5698039405    0.5065055698    0.6287518491    0.5449048325    0.2262784648
−1.4190740545  −0.7166883177    4.1476672966    0.1064317355    0.6641461068    0.7399813795    0.0031293130
−2.0233645524  −0.8880208843   −3.8414533266   −0.3222649076    0.4982433852    0.7573537759   −0.2726061566
−0.7217142988  −1.1455681173   −2.3949827521   −0.6549427515    0.1793767000    0.5657298830   −0.4677859460
 1.2569184777  −1.2389655493   −2.9988150479   −0.8039511989   −0.1233527331    0.2033003597   −0.5450830549
```

```
-1.5825097590   0.8992239464  -2.9799463334  -0.5275361451  -0.0005554527   0.0881493759   0.8449467407
 0.3185131318   0.7507260667  -2.6510951644  -0.5796510429   0.3800510422  -0.2301730359   0.6830711876
 1.6627574052   0.6449995355  -3.9628117145  -0.4546954232   0.6560307838  -0.4815616458   0.3619033906
 1.2293349601   0.8851616728   4.0726077227  -0.2211825329   0.7447937168  -0.6259998843  -0.0669682864
-0.6233564236   1.2208728662   3.5895473867   0.0465144755   0.6474946628  -0.5857509399  -0.4852658050
-2.0284746528   1.2085245669   4.8964815519   0.3464913129   0.3622051899  -0.3716219970  -0.7814398645
 0.7328133811  -0.2309745839  -4.2143017499   0.8972247986   0.4187662091  -0.1374562511  -0.0269870701
 0.2815434730  -0.0566673330   4.5335765550   0.7710826018   0.4065386142   0.4389500579  -0.2179009477
-0.8404336422   0.1611387644   4.7642064563   0.3582575939   0.2500438545   0.8438577195  -0.3115023540
-1.1358989303   0.0358894385  -3.9455686800  -0.1782628508  -0.0209102429   0.9363642119  -0.3016739638
-0.0640710550  -0.1492418508  -3.3753316411  -0.6873727038  -0.2747238096   0.6496167534  -0.1733310943
-3.1848612936  -0.5649578820   3.6153441149   0.7140782512   0.3727897241   0.2434755413   0.5402219300
-4.5345084980  -0.5604521273   2.4461966124   0.4526702465   0.5346940678   0.6474641869   0.2999700462
 3.7944895968  -0.7728371264   2.8070608391   0.0744282771   0.5346783331   0.8415548066  -0.0191054769
 2.9647892590  -1.0609239319   4.7553080461  -0.2526763232   0.3672866884   0.8351130797  -0.3222441750
 4.2571318734  -1.1359330393  -3.7729549644  -0.5871556484   0.0926783006   0.5625698417  -0.5746078231
-3.9106138908  -0.8624504858  -4.4556057012  -0.7112136094  -0.1203026573   0.1870752827  -0.6668622880
 3.5144893939   0.7399545697  -4.3518935313  -0.6610201090   0.0117012763   0.0824618703   0.7457315439
-4.8878650069   0.9771175124  -4.0233859632  -0.6594707488   0.4687309273  -0.2019571599   0.5519084660
-3.1460792272   1.2238135879   4.5736343862  -0.4916828539   0.7428853596  -0.3481142204   0.2918660707
-3.6231584497  -1.2412005475   2.7603075167  -0.1737076143   0.8326642452  -0.5134116196  -0.1135976599
 4.6995469209  -1.2281027963   2.2462639736   0.1712042524   0.7073020470  -0.4757359934  -0.4940528138
 3.2087569644   1.0774712970   3.6918193107   0.4518689025   0.4376749091  -0.3278033317  -0.7048405101
-4.2046083005   0.2933135474   4.1717871857   0.9004789306   0.4232943576  -0.0115005503   0.0991328386
-4.8981689537   0.3166941556   3.2112722836   0.7733205922   0.3053688102   0.5418685123  -0.1228969766
 4.2044034899   0.1453112079   3.5037222162   0.3538819820   0.0878925211   0.8893971791  -0.2757083701
 4.1069493014  -0.0545225507   4.6380783841  -0.1825437354  -0.1866727108   0.9046922612  -0.3366942177
-4.9160591092   0.0711685558  -4.7689399634  -0.6895818053  -0.3807816322   0.5472268501  -0.2828870039
 1.9745810915  -1.2107200772   2.4166370317   0.0216682695   0.9352531968   0.0168341380   0.3529143746
 2.1103631734   0.6579739970   3.3457752783  -0.5053109530   0.7836272048  -0.2225458090   0.2847149585
 1.6216492147   0.6566423443   1.3707769009   0.4852720520   0.8031994078   0.2683690368   0.2176230847
 2.9507009776  -0.4522661341   2.0790347613  -0.2524763293   0.8087351813   0.4830889807   0.2209709176
 0.9403239386  -0.5042883060   2.5370352541   0.2441678928   0.7974119106  -0.4885175492   0.2566454541
 0.6712892832   0.2358390057  -1.5174868582   0.2808155753   0.2153893036   0.8542927245   0.3807019852
 2.6833826332   0.2327729800  -3.5841029925   0.3140514821  -0.1934904601   0.8418943848  -0.3938869804
-2.1306712286  -0.1298037752   3.3547463428   0.3111165395  -0.3956962896   0.8360762665   0.2181912504
-4.2199316049  -0.1098825003   1.2594105222   0.2957271095   0.4139214039   0.8278224702  -0.2364836271
-3.3943881927   1.2318911358  -2.7326097940   0.0318123830   0.9813409048   0.0026191094  -0.1896078614
-2.4929500566   0.6222513535  -2.3198305955  -0.3017309989   0.8178275317  -0.4699438781  -0.1388138467
-4.4844047542   0.5720077460  -3.0003748605   0.2524664805   0.8073049416   0.4875272473  -0.2164176302
-3.1126418425  -0.5278727718  -3.5619782751  -0.4710960662   0.8187360481   0.2873077513  -0.1587262929
-3.6504596315  -0.5718451579  -1.7456480361   0.4916427813   0.8069591350  -0.2825487944  -0.1651378479
```

Configuration data for a 2×2×2 compressed approximant with $\phi = 0.8502671806$, for a compressed 8000 particle glass with $\phi = 0.7857982770$, and for a compressed 13824 particle quasicrystal with $\phi = 0.8323618782$, as well as interactive Java applets visualizing the structures, are available online at the Internet address:

`http://glotzerlab.engin.umich.edu/wiki/public/index.php/Tetrahedra`